\documentclass[12pt]{article}
\usepackage{amsfonts}
\usepackage{amsmath}
\usepackage{amssymb}
\usepackage{bm}
\usepackage{enumerate}
\usepackage{courier}
\usepackage{fullpage}
\usepackage{lineno}
\usepackage{verbatim}
\usepackage[latin1]{inputenc}
\usepackage[T1]{fontenc}         
\usepackage{ae}
\usepackage{aecompl}
\usepackage[round]{natbib}
\usepackage[english]{babel}
\usepackage[pdftex]{graphicx}   
\usepackage[bf,sf]{titlesec}
\usepackage[includemp=false, marginparwidth=1.8cm, hmargin=2.5cm, vmargin=2.8cm]{geometry}
\ifnum\pdfoutput>0			
\usepackage[pdftex, 			
pagebackref = false, 			
bookmarks = true, 			
bookmarksnumbered = false, 	
hyperindex = true,			
linktocpage = true,			
pdfpagemode = UseNone, 		
pdfstartview = Fit, 			
pdfpagelayout = OneColumn,
colorlinks = true, 			
urlcolor = blue, 				
citecolor = blue,
pdfborder = {0 0 0} 			
]{hyperref} 					
\fi
\usepackage{algorithm} 
\usepackage{algorithmicx}

\providecommand{\U}[1]{\protect\rule{.1in}{.1in}}
\newcommand{\dx}{\ensuremath{d_x}}
\newcommand{\dth}{\ensuremath{d_{\theta}}}
\newcommand{\julien}{J. Cornebise\footnote{Statistical and Applied Mathematical Sciences Institute, P.O. Box 14006, Research Triangle Park, NC 27709-4006, USA, \url{jcornebise@samsi.info}}}
\newcommand{\gareth}{G.W. Peters\footnote{School of Mathematics and Statistics, University of New South Wales, Sydney, NSW, 2052, Australia, \url{garethpeters@unsw.edu.au}}}
\newcommand{\eqsp}{\;}
\newcommand{\vect}[2]{#1 \!\!:\!\! #2}
\newcommand{\vs}{\vect{1}{S}}

\graphicspath{{./figures/}}

\title{Comments on ``Particle Markov Chain Monte Carlo'' by C. Andrieu, A. Doucet and R. Hollenstein}

\begin{document}
\author{\julien\ \ and \gareth\ \footnote{This material was based upon work supported by the National Science Foundation under Agreement No. DMS-0635449. Any opinions, findings, and conclusions or recommendations expressed in this material are those of the author(s) and do not necessarily reflect the views of the National Science Foundation.} } 
\maketitle

\begin{abstract}
We merge in this note our two discussions about the
Read Paper ``Particle Markov chain Monte Carlo'' \citep*{AndrieuDoucetHolenstein2010} presented on October 16\textsuperscript{th} 2009 at the Royal Statistical Society, appearing in the Journal of
the Royal Statistical Society Series B. We also present a more detailed version of the ABC extension. 
\end{abstract}

\section{Introduction}

The article \cite{AndrieuDoucetHolenstein2010} 
is clearly going to have significant impact on scientific 
disciplines with a strong interface with computational statistics and non-linear state space models. 
 Our comments are based on practical experience with PMCMC
implementation in latent process multifactor SDE models for commodities
\citep{PetersBriersShevchenkoDoucet2009},
wireless communications \citep{NevatPetersDoucet2009} and
population dynamics \citep{HayesHosackPeters2009}, using Rao-Blackwellised particle filters
\citep{DoucetFreitasMurphyRussel2000} and adaptive
MCMC \citep{RobertsRosenthal2009}.

\section{Generic comments}

\begin{figure}[htbp]
	\centering
	\includegraphics[width=1\textwidth]{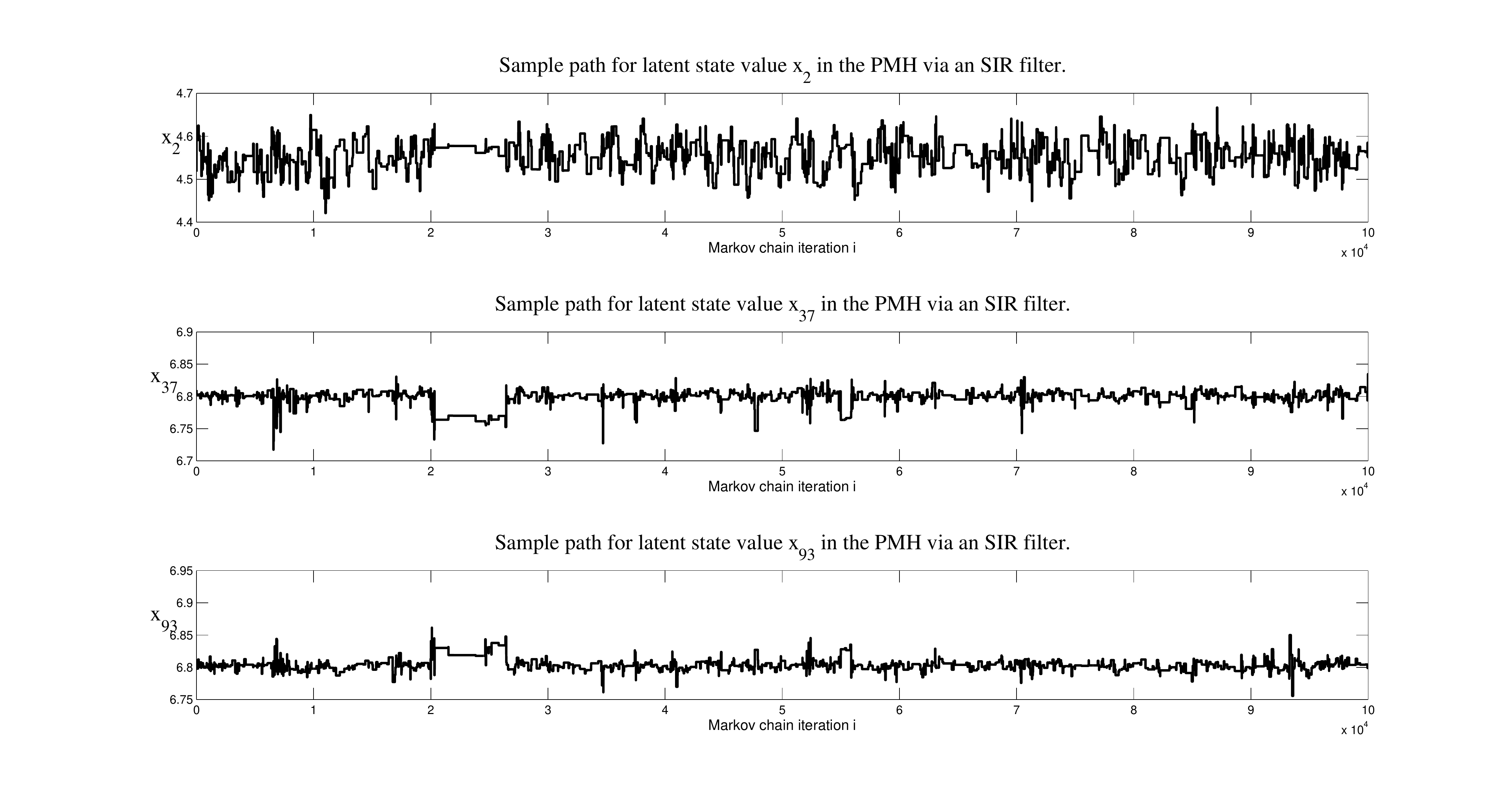}
	\vspace{-3em}
	\includegraphics[width=1\textwidth]{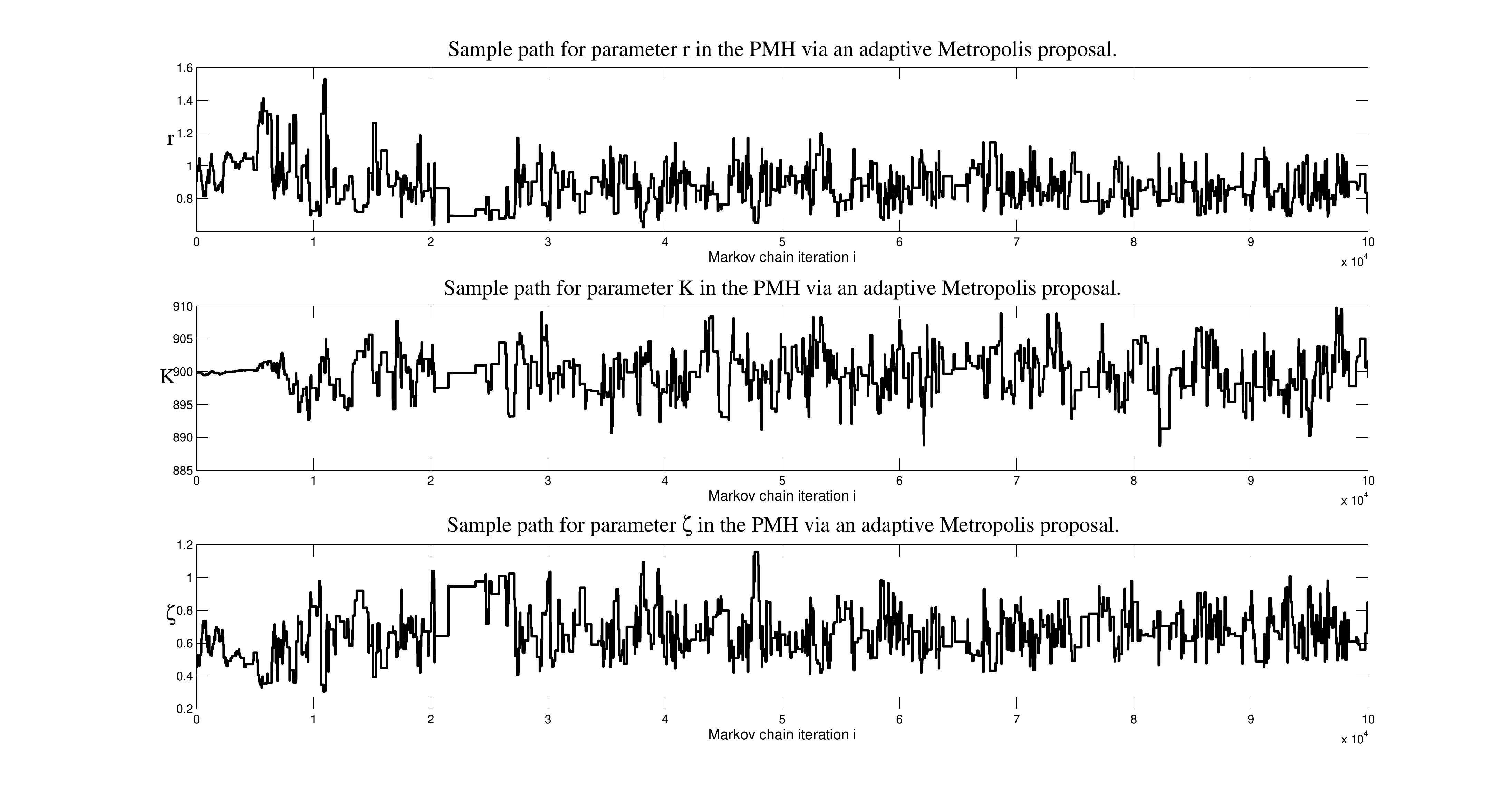} 
	\caption{Path of
	three sampled latent states $x_{2}$, $x_{37}$, $x_{93}$, and of the sampled
	parameters $\theta = (r, \zeta, L)$, over $100,000$ PMH iterations based on
	$N = 200$ particles using a simple SIR filter -- the one dimensional
	state did not call for Rao-Blackwellisation. 	Note also the effect the Adaptive MCMC proposal for $\theta$, set-up to start at iteration $5,000$, particularly visible 
	on the mixing of parameter $K$. The most noticeable property of the algorithm is the remarkable 
	mixing of the chain, in spite of the high total dimension of the sampled state: each iteration involves a
	proposal of $(X_{1:T}, \theta)$ of dimension $103$. 
}  
	\label{fig:parameters}
\end{figure}

\begin{itemize} 
	
\item From our implementations, ideal use cases consist of
	highly non-linear dynamic equations for a small dimension $\dx$ of the
	state-space, large dimension $\dth$ of the static parameter,
	and potentially large length $T$ of the time series. In our cases $\dx$ was 
	$2$ or $3$, $\dth$ up to $20$, and $T$ between $100$ and $400$.

\item	In PMH, non-adaptive MCMC proposals for $\theta$ (e.g. tuned according to 
	pre-simulation chains or burn-in iterations) would be
	costly for large $T$, and requires to keep $N$ fixed over the whole run of the Markov chain. 
	Adaptive MCMC proposals such as the Adaptive Metropolis sampler \citep{RobertsRosenthal2009}, avoid such issues and proved particularly
	relevant for large $\dth$ and $T$, as can be seen in Figure~\ref{fig:parameters}.

\item The Particle Gibbs (PG) could potentially stay frozen on a state $x_{1:T}(i)$.  
Consider a state space model with state transition function almost 
linear in $x_n$ for some range of $\theta$, from which $y_{1:T}$ is
considered to result, and strongly non-linear elsewhere. 
If the PG samples $\theta(i)$ in those regions of strong non-linearity, the particle tree would likely
coalesce on the trajectory preserved by the conditional SMC, leaving it with a high importance weight, 
maintaining $(\theta(i+1), x_{1:T}(i+1)) = (\theta(i), x_{1:T}(i))$ over several iterations. Using PMH within PG would help escape this 
region, especially using PRC and adaptive SMC kernels, outlined in another comment, to fight the 
degeneracy of the filter and the high variance of $\hat p_{\theta}(y_{1:T})$.

\end{itemize}

\section{Adaptive Sequential Monte Carlo}

Our comments on adaptive SMC relate to Particle marginal
Metropolis-Hastings (PMMH) which has acceptance probability given in Equation (13) of the read paper for proposed
state $\left(\theta^{\ast},X^{\ast}_{1:T}\right)$, relying on the estimate 
$\hat{p}_{\theta^{\ast}}\left(y_{1:T}\right) = \displaystyle \prod_{n=1}^T
\frac{1}{N}\sum_{k=1}^{N} w_n\left(x_{1:n}^{\ast,k}\right)$.
Although a small $N$ suffices to approximate the mode 
of joint path space distribution, producing a reasonable proposal for $x_{1:T}$, it results in high variance estimates
of $\hat{p}_{\theta^{\ast}}\left(y_{1:T}\right)$.
We study a population dynamics example from \citep[Model 3
excerpt]{HayesHosackPeters2009}, involving a log-transformed theta-logistic state space
model, see \cite[Equation 3(a), 3(b)]{wang2007latent} for parameter settings.
PMCMC performance depends on the trade-off between degeneracy of the filter,
$N$, and design of the SMC mutation kernel.  Regarding the latter:
\begin{itemize}
	\item	A Rao-Blackwellised filter \citep{DoucetFreitasMurphyRussel2000}
		can improve acceptance rates, \citet[see][]{NevatPetersDoucet2009}. 
	\item Adaptive mutation kernels, which in PMCMC, can be considered as
		adaptive SMC proposals, can reduce degeneracy on the path space, allowing for higher dimensional state vectors $x_n$.
		Adaption can be local (within filter) or global (sampled Markov chain history).		 
		Though currently particularly designed for ABC methods, the work of \citet{PetersFanSisson2008} incorporates into the mutation kernel of
		SMC Samplers \citep{DelmoralDoucetJasra2006} the Partial Rejection Control (PRC) mechanism of \citet{Liu2001}, which is also beneficial for 
		PMCMC. PRC adaption reduces degeneracy by rejecting a particle mutation when 
		its incremental importance weight is below a threshold $c_n$. The PRC mutation kernel
		\begin{equation}
			q_{\theta}^{\ast}(x_n|y_n,x_{n-1}) = r(c_n,x_{n-1})^{-1}
			\min 
			\left[1,W_{n-1}(x_{n-1})\frac{w_n(x_{n-1},x_n)}{c_n}\right]q_{\theta}(x_{n}|y_n,x_{n-1}),
			\label{eq:PRCkernel}
		\end{equation}
		can also be used in PMH, where $q_{\theta}(x_n|y_n,x_{n-1})$ is the standard SMC proposal, and
		\begin{equation}
			r(c_n,x_{n-1}) = \int \min \left[1, W_{n-1}(x_{n-1})
			\frac{w_n(x_{n-1},x_n)}{c_n}\right]q_{\theta}(x_n|y_n,x_{n-1}) dx_n.
			\label{eq:PRCconstant}
		\end{equation}
		As presented in \citet{PetersFanSisson2008}, algorithmic choices for $q_{\theta}^{\ast}(x_n|y_n,x_{n-1})$ can avoid 
		evaluation of $r(c_n,x_{n-1})$. \cite{Cornebise2009smcprcabc} extend this work, developing PRC for Auxiliary SMC samplers, also useful in PMH.
		Threshold $c_n$ can be set adaptively: locally either at each SMC mutation or Markov chain iteration; or globally based on chain acceptance 
		rates. Additionally, $c_n$ can be set adaptively via quantile estimates of pre-PRC incremental 
		weights, see \citet{PetersFanSisson2009}.  

	\item \citet{CornebiseMoulinesOlsson2008} state that adaptive SMC proposals 
		can be designed by minimizing function-free risk theoretic
		criteria such as Kullback-Leibler divergence between a joint
		proposal in a parametric family and a joint target. \citet[Chapter 5]{Cornebise2009} and \citet{CornebiseMoulinesOlsson2009}
		use a mixture of experts, adapting kernels of a mixture on distinct regions of the state-space separated by a softmax partition.
		These results extend to PMCMC settings.
\end{itemize}

\begin{figure}[htbp]
	\centering
	\includegraphics[width=1\textwidth]{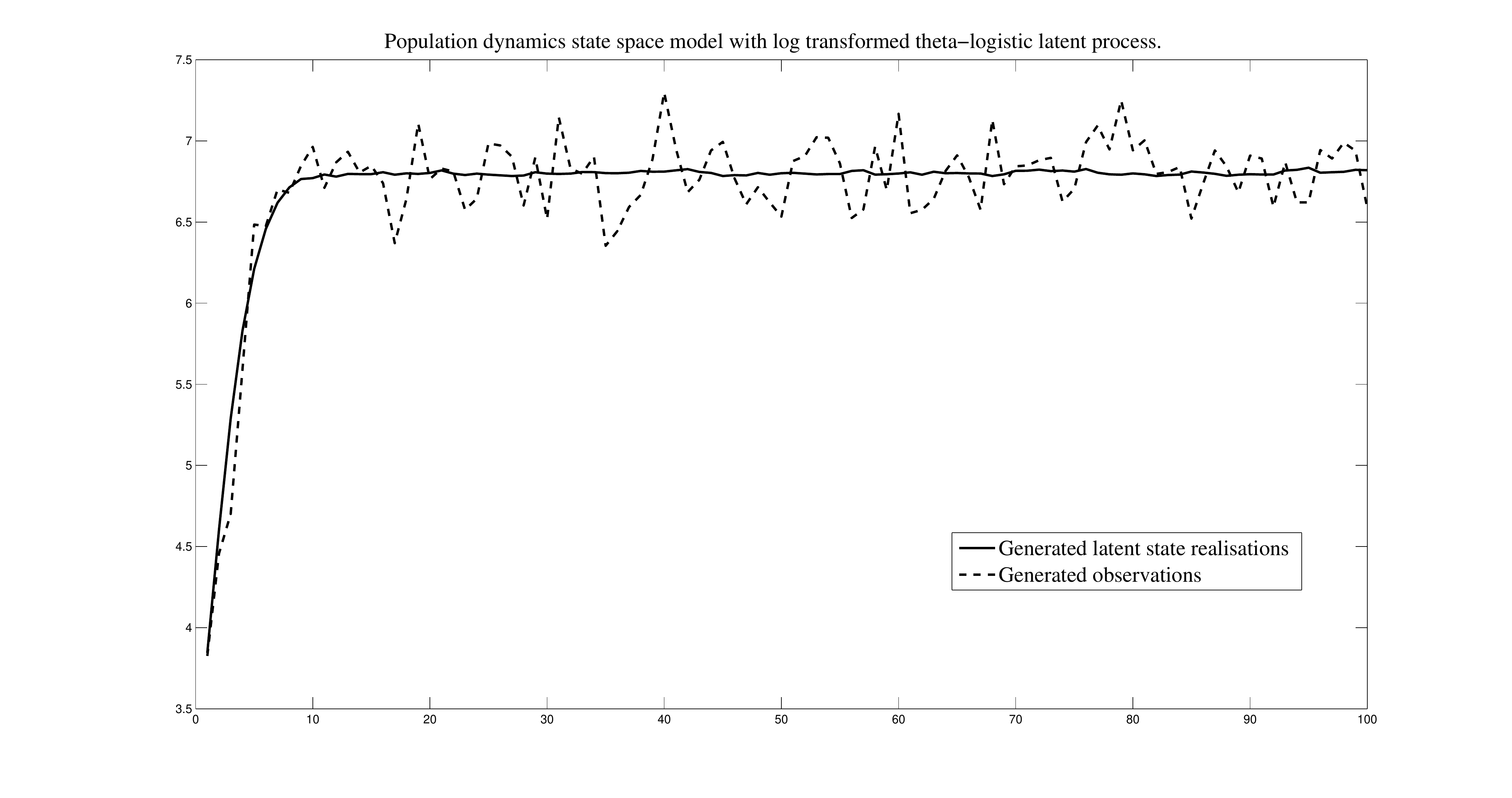}
	\caption{Sequence of simulated states and observations for the population dynamic log-transformed theta logistic model from \cite{wang2007latent}, with 
	static parameter $\theta = (r, \zeta, K)$ under constraints $K > 0$, $r < 2.69$, $\zeta \in \mathbb{R}$.
	State transition is $f_{\theta}(x_n|x_{n-1}) = \mathcal{N}\left(x_{n}; x_{n-1} + r \left( 1 - (\exp(x_{t-1})/K)^{\zeta}\right), 0.01\right)$, and local likelihood is
	$g_{\theta}(y_n|x_n) = \mathcal{N}\left( y_n ; x_n, 0.04  \right)$, for $T = 100$ timesteps.}  
	\label{fig:observations_states}
\end{figure}

\begin{figure}[htbp]
	\centering
	\includegraphics[width=1\textwidth]{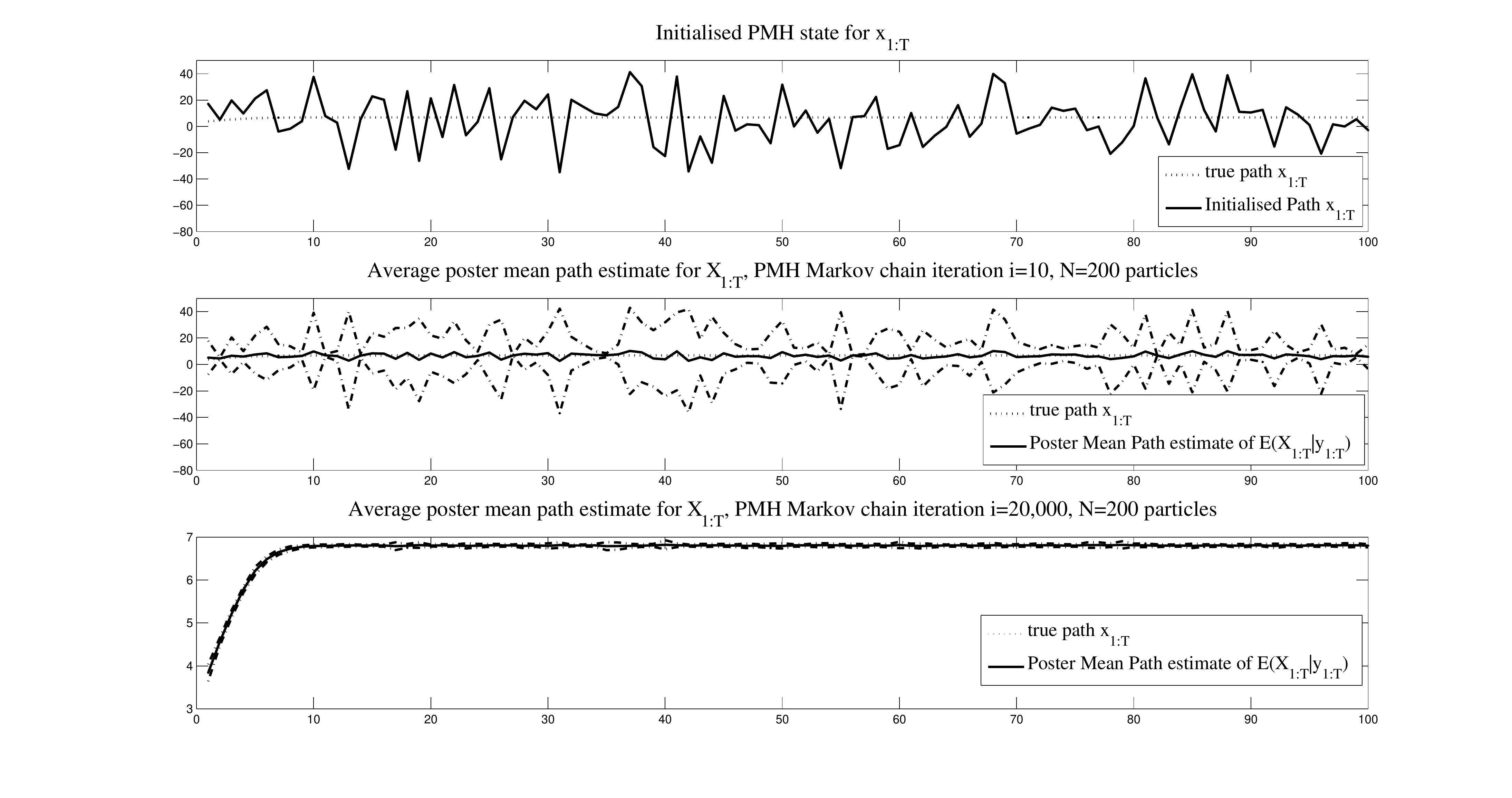}
	\caption{Convergence of the distribution of the path of latent states $x_{1:T}$. Note the change of vertical scale. Initializing PMH on a 
	very unlikely initial path does not prevent the MMSE estimate of the latent states converging: as few as $10$ PMH iterations already begins to concentrate the 
	sampled paths around the true path -- assumed here to be close to the mode of the posterior distribution thanks to the small observation noise --, with very satisfactory results
	after $20,000$ iterations.}
	\label{fig:trajectory}
\end{figure}

\section{Approximate Bayesian Computation and PMCMC}

For intractable joint likelihood $p_{\theta}(y_{1:T}|x_{1:T})$, we could design a
SMC-ABC algorithm \citep[see e.g.][Chapter 1]{PetersFanSisson2008,Ratmann2010} for a fixed ABC tolerance $\epsilon$, using the 
approximations 
	\begin{multline*}
		\hat{p}^{ABC}_{\theta}\left(y_{1:T}\right) :=
		\frac{1}{N}\sum_{k=1}^{N}
		\frac{\frac{1}{S}\sum_{s=1}^S \mathbb{I}\left(\rho(y^{k}_{1}(s),y_{1}) <
		\epsilon\right) \mu_{\theta}(x^k_{1})
		}{q_{\theta}(x^{k}_{1}|y_1)} \\
		\times
		\prod_{n=2}^T
		\left(\frac{1}{N}\sum_{k=1}^{N}
		\frac{\frac{1}{S}\sum_{s=1}^S \mathbb{I}\left(\rho(y^{k}_{n}(s),y_{n}) <
		\epsilon\right) f_{\theta}(x^k_{n}|x^{A^k_{n-1}}_{n-1})
		}{q_{\theta}(x^{k}_{n}|y_n,x^{A^k_{n-1}}_{n-1})}\right)
	\end{multline*}
or	
	\begin{multline*}
		\hat{p}^{ABC}_{\theta}\left(y_{1:T}\right) := 
		\frac{1}{N}\sum_{k=1}^{N}
		\frac{\frac{1}{S}\sum_{s=1}^S 
		\mathcal{N}\left(y^{k}_{1}(s);y_{1},\epsilon^2\right)
	 \mu_{\theta}(x^k_{1})
		}{q_{\theta}(x^{k}_{1}|y_1)} \\
		\times
		\prod_{n=2}^T
		\left(\frac{1}{N}\sum_{k=1}^{N}
		\frac{\frac{1}{S}\sum_{s=1}^S \mathcal{N}\left(y^{k}_{n}(s);y_{n},\epsilon^2\right) f_{\theta}(x^k_{n}|x^{A^k_{n-1}}_{n-1})
		}{q_{\theta}(x^{k}_{n}|y_{n},x^{A^k_{n-1}}_{n-1})}\right)
	\end{multline*}
	with $\rho$ a distance on the observation space and $y^{k}_{n}(s) \sim g_{\theta}(\cdot|x^k_n)$ simulated observations.
	Additional degeneracy on the path space induced by ABC approximation should be controlled, e.g. with PRC \citep{PetersFanSisson2008}, see
	Equation~\eqref{eq:PRCkernel}. More details on this algorithm are available in Appendix~\ref{sec:appendixABC}, which is not contained
	in our comment to JRSSB due to space restrictions.

\appendix
\section{Algorithmic details of ABC filtering within PMCMC}
\label{sec:appendixABC}

\begin{algorithm}[htbp]
	\caption{SMC-ABC-PRC filtering algorithm targeting $p_{\theta}(x_{1:T} | y_{1:T})$ as required in Step 2(b) of the PMMH of \cite[Section 2.4.2]{AndrieuDoucetHolenstein2010}. Replaces the SMC algorithm presented in \cite[Section 2.2.1]{AndrieuDoucetHolenstein2010}. Approximation $g_{\theta}^{ABC}$ is defined in Equation~\eqref{eq:abc:loclike} and function $r$ in Equation~\eqref{eq:PRCconstant}.}

	\begin{enumerate}[\emph{{Step }}1:]
		\item Initialize $\epsilon$ and $c_1$
		\item At time $n = 1$,
			\begin{enumerate}[(a)]
				\item for $k=1, \ldots, N$
					\begin{enumerate}[(i)]
						\item sample $X^{k}_1 \sim q_{\theta}(\cdot|y_1)$
						\item sample $Y_1^k(s) \sim g_{\theta}(\cdot|X^{k}_1)$ for $s = 1, \ldots, S$
						\item compute the incremental weight
					\begin{equation*}
						\tilde w_1( X^k_1) := 
						\frac{\mu_{\theta}(X^k_1) g_{\theta}^{ABC}(y_1|X^k_1,Y_1^k(\vs),\epsilon)}
						{q_{\theta}(X^k_1|y_1)}\eqsp,
					\end{equation*}
				\item with probability $1 - p^k_1 = 1 - \min\{1,  \tilde w_1( X^k_{1})/c_1\}$, reject $X^k_1$ and go to (i)
				\item otherwise, accept $X^k_1$ and set 
					\begin{equation*}
						w_1( X^k_{1}) = \tilde w_1( X^k_{1}) r(c_1)/ p^k_1
					\end{equation*}
			\end{enumerate}
		\item normalise the weights $W_1^k := w_1(X^k_1)/\sum_{m=1}^N w_1(X^m_1)$.
			\end{enumerate}
		\item At times $n = 2, \ldots, T$,
			\begin{enumerate}[(a)]
				\item possibly adapt $c_n$ online
				\item for $k=1, \ldots, N$

					\begin{enumerate}[(i)]
				\item sample $A^k_{n-1} \sim \mathcal{F}(\cdot|\mathbf{W}_{n-1})$,
				\item sample $X^{k}_n \sim q_{\theta}(\cdot|y_n,X^{A^k_{n-1}}_{n-1})$ 
					and set $X^{k}_{1:n} = (X^{A^k_{n-1}}_{n-1},X^{k}_n)$, and
				\item sample $Y_n^k(s) \sim g_{\theta}(\cdot|X^{k}_n)$ for $s = 1, \ldots, S$
				\item compute the incremental weight
					\begin{equation*}
						\tilde w_n( X^k_{1:n}) := 
						\frac{f_{\theta}(X^k_n|X^{A^k_{n-1}}_{n-1}) g_{\theta}^{ABC}(y_n|X^k_n,Y_n^k(\vs),\epsilon)}
						{q_{\theta}(X^k_n|y_n,X^{A^k_{n-1}}_{n-1})}\eqsp,
					\end{equation*}
				\item with probability $1 - p^k_n = 1 - \min\{1,  \tilde w_n( X^k_{1:n})/c_n\}$, reject $X^k_n$ and go to (ii)
				\item otherwise, accept $X^k_n$ and set 
					\begin{equation*}
						w_n( X^k_{1:n}) = \tilde w_n( X^k_{1:n}) r(c_n, X^{A^k_{n-1}}_{n-1})/ p^k_n
					\end{equation*}
			\end{enumerate}
				\item normalise the weights $W_n^k := w_n(X^k_{1:n}) / \sum_{m=1}^N w_n(X^m_{1:n})$.
			\end{enumerate}
	\end{enumerate}
\label{alg:ABC}
\end{algorithm}

Here we expand on the comment we made above in which we approximated the
local likelihood  $g_{\theta}(y_n|x_n)$ 
of the SMC-based filtering part of the PMCMC algorithm.
by the ABC approximation 
\begin{equation}
	g_{\theta}^{ABC}(y_n|x_n,y_n(\vs),\epsilon) := \frac{1}{S}\sum_{s=1}^S 
	\pi_{\theta}\left( y_n(s) |x_n, y_n, \epsilon  \right)
	\label{eq:abc:loclike}
\end{equation}
where possible choices for $\pi_{\theta}$ are
\begin{equation*}
	\pi^{\mathbb{I}}_{\theta}\left( y_n(s) |x_n, y_n, \epsilon  \right) 
 := \mathbb{I}\left(\rho(y_{n}(s),y_{n}) <
\epsilon\right) \text { or } 
\pi^{\mathcal{N}}_{\theta}\left( y_n(s) |x_n, y_n, \epsilon  \right) 
 :=\mathcal{N}\left(y_{n}(s);y_{n},\epsilon^2\right)
\end{equation*}
with $\rho$ a distance on the observation space and $y_{n}(s) \sim g_{\theta}(\cdot|x_n)$ simulated observations
 -- assumed here to be univariate for sake of brevity, but generalisation to multivariate setting and summary statistics is straightforward. 

We note it is critical in the filtering context to ensure that the particle
system under approximation does not collapse into uniformly null
incremental weights $w_n\left( X^k_{1:n} \right) = 0$ which may occur for
$\pi^{\mathbb{I}}_{\theta}$ at any stage of the filtering during each PMCMC
iteration, especially for small tollerances $\epsilon$.  The PRC mutation
kernel $q^{\ast}_{\theta}(x_n|y_n,x_{n-1})$ defined in
Equation~\eqref{eq:PRCkernel} is critical to overcome both this collapse and
the additional degeneracy on the path space introduced by the ABC
approximation. The algorithm presented in \citet[Sections 3.4 and
3.5.1]{McKinleyCookDeardon2009} is a special case of the SMC samplers PRC-ABC
algorithm of \cite{PetersFanSisson2008} in which the PRC rejection
threshold $c_n = 0$, the mutation kernel is global and resampling is
performed at each stage of the filter, which avoids the computation of the
normalizing constant $r(c_n,x_{n-1})$ defined in
Equation~\eqref{eq:PRCconstant}. We further note that the work of
\citet{Cornebise2009} casts the SMC sampler PRC algorithm \citep{PetersFanSisson2008} 
with rejection of the ancestor index $A^k_{n-1}$ -- here in Step 3.(a).(v) -- 
into an Auxiliary SMC sampler framework.
The combination of these two concepts recovers a generalized version of 
\cite{McKinleyCookDeardon2009}, see Algorithm~\ref{alg:ABC}, which has 
advantage in the PMCMC setting of allowing for 
adaptation of the threshold $c_n$.

\bibliographystyle{chicago}
\bibliography{biblio}

\end{document}